# Simple Stabilized Radio-Frequency Transfer with Optical Phase Actuation

DAVID R. GOZZARD,[1,2,*] SASCHA W. SCHEDIWY,[2,1] RICHARD WHITAKER[3], KEITH GRAINGE[4]

[1]School of Physics and Astrophysics, University of Western Australia, Perth, WA 6009, Australia
[2]International Centre for Radio Astronomy Research, University of Western Australia, Perth, WA 6009, Australia
[3]Jodrell Bank Observatory, University of Manchester, M13 9PL, United Kingdom
[4]Jodrell Bank Centre for Astrophysics, Alan Turing Building, School of Physics & Astronomy, The University of Manchester, Oxford Road, Manchester M13 9PL, UK
*Corresponding author: david.gozzard@research.uwa.edu.au

**We describe and experimentally evaluate a stabilized radio-frequency transfer technique that employs optical phase sensing and optical phase actuation. This technique can be achieved by modifying existing stabilized optical frequency equipment and also exhibits advantages over previous stabilized radio-frequency transfer techniques in terms of size and complexity. We demonstrate the stabilized transfer of a 160 MHz signal over an 166 km fiber optical link, achieving an Allan deviation of 9.7×10⁻¹² Hz/Hz at 1 s of integration, and 3.9×10⁻¹⁴ Hz/Hz at 1000 s. This technique is being considered for application to the Square Kilometre Array SKA1-low radio telescope.**

Fiber-optic frequency transfer networks, such as the European NEAT-FT [1 - 3] and the Beijing regional time and frequency network [4, 5], are now being rolled-out on national and international scales, driven by cutting-edge research in metrology and other physical sciences including the distant comparison of optical atomic clocks, high-precision remote spectroscopy, radio astronomy, geodesy, and tests of fundamental physics [1, 6, 7]. The NEAT-FT network, and other long-distance frequency comparison experiments [2, 8, 9], have typically transmitted stabilized optical-frequency signals because of their superior fractional frequency stability over optically-disseminated radio-frequency (RF) and microwave (MW) signals [1, 6]. However, many applications of stabilized frequency transfer in science, commerce, and industry require RF transmissions [7], and will benefit even at the expense of inferior stability performance compared to optical frequency transfer. These applications cannot be interfaced directly with optical frequencies, and while optical-to-RF conversion methods exist [10, 11, 12], the required equipment remains expensive and complex.

Stabilized RF and MW dissemination systems typically require bulky group-delay actuators (thermally controlled fiber spools) [13, 14] or a separate remote-site transmission system operating in a phase-locked-loop [15, 16], and are not compatible with existing optical frequency transfer infrastructure such as that demonstrated in [2], [8] and [9].

In this paper, we present a modification of the technique reported in [17] to create a simple and compact stabilized RF transfer system that uses only the optical and electronic components used in standard stabilized optical transfer systems [18], to enable existing optical transfer infrastructure to readily be converted to provide stabilized RF transfer.

As shown in Figure 1, an optical signal with frequency $\nu_L$, is generated by a laser located at the **Local Site**. Just as is the case in standard stabilized optical transfer techniques [18], the optical signal enters an imbalanced Michaelson interferometer (MI) via an optical isolator (to prevent reflections returning to the laser). The short arm of the MI provides the physical reference for the optical phase sensing. The optical reference signal $\nu_{\text{ref}}$ at the photodetector is

$$\nu_{\text{ref}} = \nu_L + \frac{1}{2\pi}\left(2\Delta\dot{\phi}_{\text{MI}}\right), \tag{1}$$

where $\Delta\phi_{\text{MI}}$ is the undesirable phase noise picked up by the optical signals passing through the MI reference arm.

In the long arm of the MI the optical signal is then split into two arms of a Mach-Zehnder interferometer (MZI), each arm of which contains an acousto-optic modulator (AOM). In the case shown in Figure 1, the bottom arm of the MZI contains the servo AOM, which has a nominal frequency $\nu_{A-\text{srv}}$, and also applies the frequency correction $\Delta\nu_{A-\text{srv}}$. The frequency fluctuations due to optical path length changes in the bottom arm of the MZI are represented by $\frac{1}{2\pi}\Delta\dot{\phi}_{\text{MZI},2}$. The top arm of the MZI contains the local 'anti-reflection' AOM, with frequency $\nu_{A-\text{lar}}$. The frequency fluctuations in this arm of the MZI are given by $\frac{1}{2\pi}\Delta\dot{\phi}_{\text{MZI},1}$. (Note — the anti-reflection AOM is not essential for this technique. It serves to provide reflection mitigation as well as increase the frequency separation.) A combination of up- and down-shifting AOMs gives the greatest RF separation, but any unique combination can be used.

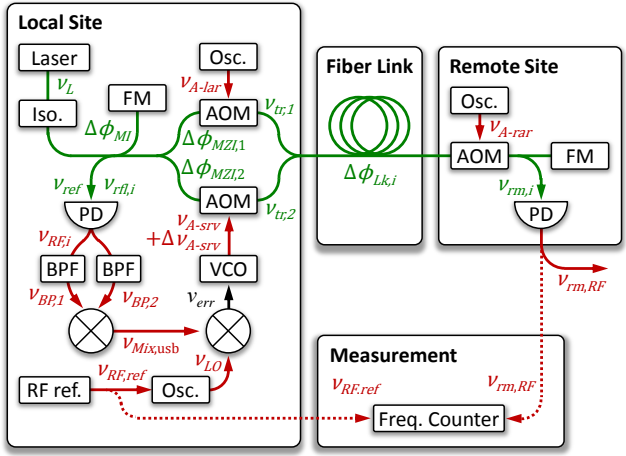

**Fig. 1.** Schematic diagram of our stabilized radio-frequency transfer technique. Optical-frequency signals are shown in green; radio-frequency signals in red; and the error signal in black. AOM acousto-optic modulator; Osc. Oscillator; Iso. optical isolator; FM Faraday mirror; PD photodetector; VCO voltage-controlled oscillator; BPF band-pass filter; and RF ref. radio-frequency reference.

At the output of the MZI, the two optical signals to be transmitted are now:

$$\nu_{tr,1} = \nu_L + \nu_{A-\text{lar}} + \frac{1}{2\pi}\Delta\dot{\phi}_{MZI,1}, \text{ and} \tag{2}$$

$$\nu_{tr,2} = \nu_L + (1+\Delta)\nu_{A-\text{srv}} + \frac{1}{2\pi}\Delta\dot{\phi}_{MZI,2}. \tag{3}$$

As the two optical signals pass through the link, they pick-up frequency fluctuations, $\Delta\dot{\phi}_{Lk,i}$, due to optical path length changes in the link that are unique to their specific transmitted frequency.

At the remote site, the two optical signals pass through a remote anti-reflection AOM (again, this AOM is not essential for the technique, but allows rejection of unwanted reflected signals) to give:

$$\nu_{rm,1} = \nu_L + \nu_{A-\text{lar}} + \nu_{A-\text{rar}} + \frac{1}{2\pi}(\Delta\dot{\phi}_{MZI,1} + \Delta\dot{\phi}_{Lk,1}), \tag{4}$$

and

$$\nu_{rm,2} = \nu_L + (1+\Delta)\nu_{A-\text{srv}} + \nu_{A-\text{rar}} + \frac{1}{2\pi}(\Delta\dot{\phi}_{MZI,2} + \Delta\dot{\phi}_{Lk,2}). \tag{5}$$

At the **Remote Site**, the signal is split with one part going to a photodetector. The electronic signal $\nu_{rm,e}$ from the beat of $\nu_{rm,1}$ and $\nu_{rm,2}$ is

$$\nu_{rm,e} = \frac{1}{2\pi}(\Delta\dot{\phi}_{MZI,2} - \Delta\dot{\phi}_{MZI,1} + \Delta\dot{\phi}_{Lk,2} - \Delta\dot{\phi}_{Lk,1}) + (1+\Delta)\nu_{A-\text{srv}} - \nu_{A-\text{lar}}. \tag{6}$$

A mirror reflects the two optical signals back through the link to the **Local Site** where they then pass back though the MZI. The returning reflected optical signals reaching the photodetector for the MI are then:

$$\nu_{rfl,1} = \nu_L + 2\left(\nu_{A-\text{lar}} + \nu_{A-\text{rar}} + \frac{1}{2\pi}(\Delta\dot{\phi}_{MZI,1} + \Delta\dot{\phi}_{Lk,1})\right), \tag{7}$$

$$\nu_{rfl,2} = \nu_L + 2\left((1+\Delta)\nu_{A-\text{srv}} + \nu_{A-\text{rar}} + \frac{1}{2\pi}(\Delta\dot{\phi}_{MZI,2} + \Delta\dot{\phi}_{Lk,2})\right), \tag{8}$$

and

$$\nu_{rfl,3j} = \nu_L + (1+\Delta)\nu_{A-\text{srv}} + \nu_{A-\text{lar}} + 2\nu_{A-\text{rar}} + 2\Delta\dot{\phi}_{Lk,j} + \frac{1}{2\pi}(\Delta\dot{\phi}_{MZI,1} + \Delta\dot{\phi}_{MZI,2}), \tag{9}$$

where $j$ is 1 or 2 corresponding to the signals on the link. The three signals are at unique frequencies as long as $\nu_{A-\text{srv}}$ does not equal $\nu_{A-\text{lar}}$. At the photodetector these optical frequencies mix with $\nu_{ref}$ to give the following RF signals.

$$\nu_{RF,1} = 2\left(\nu_{A-\text{lar}} + \nu_{A-\text{rar}} + \frac{1}{2\pi}(\Delta\dot{\phi}_{MZI,1} + \Delta\dot{\phi}_{Lk,1} - \Delta\dot{\phi}_{MI})\right), \tag{10}$$

$$\nu_{RF,2} = 2\left((1+\Delta)\nu_{A-\text{srv}} + \nu_{A-\text{rar}} + \frac{1}{2\pi}(\Delta\dot{\phi}_{MZI,2} + \Delta\dot{\phi}_{Lk,2} - \Delta\dot{\phi}_{MI})\right), \tag{11}$$

and

$$\nu_{RF,3j} = (1+\Delta)\nu_{A-\text{srv}} + \frac{1}{2\pi}(\Delta\dot{\phi}_{MZI,1} + \Delta\dot{\phi}_{MZI,2} + 2\Delta\dot{\phi}_{Lk,j}) + \nu_{A-\text{lar}} + 2\nu_{A-\text{rar}}, \tag{12}$$

as well as intermodulation signals. In the electronic domain, the signals are split and bandpass filtered. The RF bandpass filter values are set to $\nu_{BP,1} = 2(\nu_{A-\text{lar}} + \nu_{A-\text{rar}})$ and $\nu_{BP,2} = 2(\nu_{A-\text{srv}} + \nu_{A-\text{rar}})$.

The bandpass filters eliminate $\nu_{RF,3j}$, the intermodulation signals, and the opposing RF signal. The filtered signals are then mixed together, producing the following upper- and lower-sideband frequency products:

$$\nu_{Mix,1} = 2\left((1+\Delta)\nu_{A-\text{srv}} + \nu_{A-\text{lar}} + \frac{1}{2\pi}(\Delta\dot{\phi}_{MZI,1} + \Delta\dot{\phi}_{MZI,2} + \Delta\dot{\phi}_{Lk,1} + \Delta\dot{\phi}_{Lk,2} - 2\Delta\dot{\phi}_{MI})\right), \tag{13}$$

and

$$\nu_{Mix,2} = 2\left((1+\Delta)\nu_{A-\text{srv}} - \nu_{A-\text{lar}} + \frac{1}{2\pi}(\Delta\dot{\phi}_{MZI,1} - \Delta\dot{\phi}_{MZI,2} + \Delta\dot{\phi}_{Lk,1} - \Delta\dot{\phi}_{Lk,2})\right). \tag{14}$$

Note that in $\nu_{Mix,2}$ the frequency perturbation $\Delta\dot{\phi}_{MI}$ cancels out. A bandpass filter with the center frequency at $2 \times (\nu_{A-\text{srv}} - \nu_{A-\text{lar}})$ is used to reject $\nu_{Mix,1}$, before mixing $\nu_{Mix,2}$ with the servo local oscillator $\nu_{LO}$ also set at $\nu_{LO} = 2(\nu_{A-\text{srv}} - \nu_{A-\text{lar}})$ to produce an error signal of

$$\nu_{err} = 2\left(\nu_{A-\text{srv}} + \frac{1}{2\pi}(\Delta\dot{\phi}_{MZI,2} - \Delta\dot{\phi}_{MZI,1} + \Delta\dot{\phi}_{Lk,2} - \Delta\dot{\phi}_{Lk,1})\right). \tag{15}$$

When the servo is engaged, the error signal is driven to zero, $\nu_{err} = 0$, so:

$$\nu_{A-\text{srv}} = -\frac{1}{2\pi}(\Delta\dot{\phi}_{MZI,2} - \Delta\dot{\phi}_{MZI,1} + \Delta\dot{\phi}_{Lk,2} - \Delta\dot{\phi}_{Lk,1}). \tag{16}$$

Substituting this into equation 6 gives:

$$\nu_{\text{rm,e}*} = \nu_{A-\text{srv}} - \nu_{A-\text{lar}}, \quad (17)$$

where $\nu_{\text{rm,e}*}$ is the electronic remote signal with the servo engaged.

We describe an experiment using 160 MHz RF transfer, with all optical elements fiberized. An NKT Photonics Koheras BASIK X15 laser (spectral linewidth <100 Hz) situated at the **Local Site**, and operating at a wavelength of 1552 nm, produced a laser signal $\nu_L$ = 193 THz. The AOMs were Gooch & Housego with $\nu_{A-\text{srv}}$ = +75 MHz, $\nu_{A-\text{lar}}$ = −85 MHz, and $\nu_{A-rar}$ = 40 MHz. All fiber in the **Local Site** was polarization maintaining to ensure the optical power output of the MZI remains maximized. The signal was transmitted through AARNet-managed metropolitan optical fiber networks 166 km in length, with two IDIL Fibres Optiques bi-directional optical amplifiers used to boost the signal strength as required.

Menlo FPD-510 photodetectors were used for the optical-to-electronic conversion in both the **Local Site** and **Remote Site**. Faraday mirrors were used at the ends of the two arms of the MI to ensure the signals mixing at the transmitter side photodetector were aligned in polarization.

The combination of AOM frequencies resulted in the following electronic signals within the servo electronics; $\nu_{\text{RF},1}$ = 90 MHz, $\nu_{\text{RF},2}$ = 230 MHz, and $\nu_{\text{RF},3}$ = 70 MHz. The mixer frequencies are $\nu_{\text{Mix},1}$ = 140 MHz and $\nu_{\text{Mix},2}$ = 320 MHz. After filtering, $\nu_{\text{Mix},2}$ was mixed with a 320 MHz LO frequency.

Both the **Local Site** and **Remote Site** were co-located in the same laboratory, permitting an independent measure of the transfer stability. The 160 MHz received signal was terminated on a Π-type Microsemi 5125A phase noise test set to produce an Allan deviation estimate of the fractional frequency stability.

Figure 2 shows the Allan deviation measured for the stabilized (filled blue circles, solid lines) and unstabilized (open blue circles, dashed lines) 160 MHz signal transmitted over the 166 km link.

Figure 2 shows that the stabilization system produces an Allan deviation of 9.7×10⁻¹² Hz/Hz at 1 s of integration, dropping to 3.9×10⁻¹⁴ Hz/Hz at 1000 s integration time.

We have described and experimentally demonstrated the function of a stabilized radio-frequency transfer technique. This technique exploits several advantages of optical phase-sensing and optical phase-actuation over other radio- and microwave-frequency transfer techniques.

The use of AOMs for optical phase-actuation allows the **Local Site** transmitter units to have a more compact construction than systems using fiber stretchers or thermal spools. AOMs also provide an infinite feedback range, avoiding any potential need for integrator-reset or range-limit monitoring circuits. The rapid response of AOMs allows for better optimization of the servo gains for short links where the feedback bandwidth is limited by electronic delay rather than the light round-trip time.

This technique actively suppresses phase noise introduced in the **Local Site** MZI as well as the transmission link. Along with the simplicity of the remote site optical and electronic components, this makes the system very resistant to the effects of environmental perturbations (such as temperature variation or vibration) on the system hardware.

By using AOMs to generate shifts of the optical frequency at the local and remote sites, this technique avoids the need to use additional lasers at the remote sites to circumvent the effects of unwanted reflections on the transmission link. This reduces the complexity and cost of the system. In addition, on links where unwanted reflections are minimal, it is possible to remove an anti-reflection AOM from the stabilized transfer system (either the remote AOM, or the local AOM), to further reduce the cost and complexity. The local and the remote anti-reflection AOMs mitigate unwanted reflections by generating unique RF frequencies at the photodetectors. Sensible choices of AOM frequencies allow this to happen without overlapping with signals of interest.

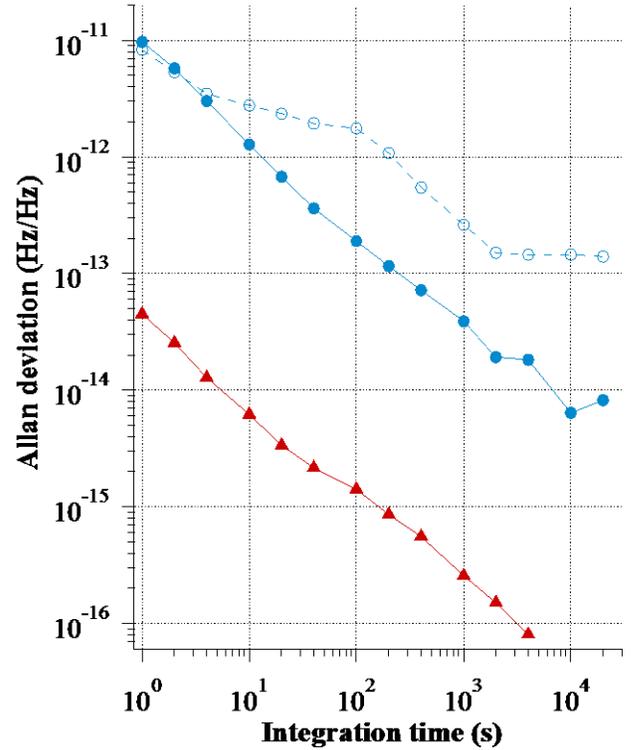

**Fig. 2.** Allan deviation of 160 MHz transfer over 166 km of metropolitan fiber. Stabilization servo engaged — filled blue circles, solid line. Stabilization disengaged — open blue circles, dashed line. Electronic noise floor of the phase-noise test set — filled red triangles, solid line.

The use of Faraday mirrors at the end of the MI arms ensures that the maximum signal is available at the photodetectors without the need for any initial polarization alignment or ongoing polarization control or polarization scrambling.

Because the RF signal being transmitted is the product of only two optical signals, and not three as is the case in standard intensity modulation, the system avoids potential problems of signal fading that is caused by the destructive interference of the modulation sidebands at specific intervals along the link. This also means that the transmission frequency can be varied arbitrarily without needing to consider the link length. The techniques employed mean that the system does not require the use of specialized fiber such as dispersion compensating fiber or polarization maintaining fiber. Furthermore, the technique is compatible with data transmission on the same link, allowing the system to make use of active telecommunications links if the non-bi-directional components of the link are bypassed as in [6] and [19]. Bi-directional optical amplifiers can be used to simply extend the range of the system without the use of signal regeneration stations at certain points along the link.

The technique of splitting and frequency shifting the signal from a single laser source is also considerably simpler than offset-locking a master and slave laser, however, the range of possible transmission frequencies is limited to the RF domain (that is, < 1 GHz) by the practical limitations of existing AOM technology. The largest frequency shifts produced by commercially available AOMs at the time of publication are +/-110 MHz. Including additional passive AOMs can further increase the transmission frequency. However, due to the optical power loss of the AOMs, it is not currently practical to achieve stabilized MW transmission (> 1 GHz) using the technique presented here. Stabilized MW transmissions using optical phase-sensing and actuation require the more extensive modifications presented in [17].

While testing other transmission frequencies, we found that when the system is configured correctly — that is, optical and electronic signal levels are optimized and adequate filters are employed — the absolute frequency stabilities of the transmitted signals are very similar (within a factor of about 2) regardless of the transmission frequency. This observation has been found to hold true for the stabilized transmission of a frequency up to 50 times higher (the 8 GHz frequency transfer presented in [17]). Therefore, the fractional frequency stability can be improved by transmitting higher transmission frequencies.

This stabilized RF transfer technique is one of two being considered for selection as the phase-synchronization system for the Square Kilometre Array SKA1-low radio telescope [20, 21].

**Funding.** University of Manchester; University of Western Australia (UWA).

**Acknowledgment**. The authors wish to thank *AARNet* for the provision of light-level access to their fiber network infrastructure. This paper describes work being carried out for the SKA Signal and Data Transport (SaDT) consortium as part of the Square Kilometre Array (SKA) project. The SKA project is an international effort to build the world's largest radio telescope, led by SKA Organisation with the support of 10 member countries. This work was supported by funds from the University of Manchester and University of Western Australia.